\documentclass[aps,prd,twocolumn,showpacs,superscriptaddress]{revtex4} 

\usepackage{graphicx}
\usepackage{multirow}
\usepackage{amsmath}

\font\FermiSmallfont=cmssq8 scaled 1200

\def\LANLppthead#1{
\null 
\begin{center}\vskip -1.0truein{\hbox to 7.5truein {
\hfill
\vbox to 1in {\vfill \FermiSmallfont
              \hbox{#1}
              \vfill}
}}\vskip-0.0truein\end{center}}

\begin{document}

\title{Heavy sterile neutrinos, entropy and relativistic energy production, and the relic neutrino background}
\author{George M. Fuller}
\affiliation{Department of Physics, University of California, San Diego, La Jolla, CA 92093-0319}
\author{Chad T. Kishimoto}
\affiliation{Department of Physics and Astronomy, University of California, Los Angeles, California 90095}
\author{Alexander Kusenko}
\affiliation{Department of Physics and Astronomy, University of California, Los Angeles, California 90095}
\affiliation{Institute for the Physics and Mathematics of the Universe, University of Tokyo, Kashiwa, Chiba 277-8568, Japan}


\date{\today}

\begin{abstract}
We explore the implications of the existence of heavy neutral fermions ({\it e.g.,} sterile neutrinos) for the thermal history of the early universe. In particular, we consider sterile neutrinos with rest masses in the $100\,{\rm MeV}$ to $500\,{\rm MeV}$ range, with couplings to ordinary active neutrinos large enough to guarantee thermal and chemical equilibrium at epochs in the early universe with temperatures $T > 1\,{\rm GeV}$, but in a range to give decay lifetimes from seconds to minutes. Such neutrinos would decouple early, with relic densities comparable to those of photons, but decay out of equilibrium, with consequent prodigious entropy generation prior to, or during, Big Bang Nucleosynthesis (BBN). Most of the ranges of sterile neutrino rest mass and lifetime considered are at odds with Cosmic Microwave Background (CMB) limits on the relativistic particle contribution to energy density ({\it e.g.,} as parameterized by $N_{\rm eff}$). However, some sterile neutrino parameters can lead to an acceptable $N_{\rm eff}$. These parameter ranges are accompanied by considerable dilution of the ordinary background relic neutrinos, possibly an adverse effect on BBN, but sometimes fall in a range which can explain measured neutrino masses in some particle physics models. A robust signature of these sterile neutrinos would be a measured $N_{\rm eff} \neq 3$ coupled with {\it no} cosmological signal for neutrino rest mass when the detection thresholds for these probes are below laboratory-established neutrino mass values, either as established by the atmospheric neutrino oscillation scale or direct measurements with, {\it e.g.,} KATRIN or neutrino-less double beta decay experiments.

\end{abstract}
\maketitle

\section{Introduction}

In this paper we show that recent Cosmic Microwave Background (CMB) limits on the proportion of energy density in the early universe contributed by particles with relativistic kinematics can provide powerful constraints on a possible sector of particle physics which is difficult and sometimes impossible to probe in the laboratory. The constraints we derive are independent of Big Bang Nucleosynthesis (BBN) considerations, though these may ultimately extend and strengthen our limits. 

Sterile neutrino effects on the conditions in BBN have been considered in Ref.s \cite{Scherrer:1984fr,Dolgov:2000fr}. Likewise, much work has been done on the effects of heavy particle decay on BBN and on particle decay-generated, post-BBN cascade nuclear reactions and nucleosynthesis \cite{Jedamzik:2009fj,Jedamzik:2006yq,Jedamzik:2004kx,Kawasaki:2005fk,Kawasaki:2005lr,Cumberbatch:2007qy,Pospelov:2007vn,Ishiwata:2010uq,Cyburt:2010rt,Ellis:2011vn,Pospelov:2011uq,Pospelov:2010fj}. Here we concentrate on a range of sterile neutrino masses and lifetimes which are, in some cases, different than what has been studied before and we concentrate on effects on the thermal neutrino background and the thermodynamics of the early universe.

The experimental and observational establishment of neutrino flavor mixing and nonzero neutrino rest masses invites speculation about new beyond-Standard-Model physics in the neutrino sector. In particular, the existence of right-handed neutral fermions (sterile neutrinos) seems at least plausible. Right-handed neutrinos with masses below the electroweak scale are called sterile neutrinos. The LEP results require only three active neutrinos with standard weak interactions, but there is no limit on the number of sterile neutrinos. There is no compelling argument for the rest mass scales of these states, with models invoking masses from the sub-eV range to the unification scale. Sterile neutrinos have been invoked as a way to engineer successful $r$-Process nucleosynthesis in neutrino-heated supernova ejecta \cite{McLaughlin:1999fk,Caldwell:2000db,Fetter:2003lr}. Sterile neutrinos with $\sim {\rm keV}$ rest masses have been studied as potential dark matter candidates~\cite{Dodelson:1994rt,Shi:1999lq,Abazajian:2001lr,Dolgov:2002ve,Abazajian:2002bh,Asaka:2005fj,Abazajian:2006qf,Shaposhnikov:2006fj,Boyanovsky:2007lr,Boyanovsky:2007fk,Shaposhnikov:2007qy,Gorbunov:2007uq,Kishimoto:2008pd,Laine:2008kx,Petraki:2008yq,Petraki:2008vn}, and as enablers of the formation of the first stars \cite{Biermann:2006mz,Mapelli:2006gf,Stasielak:2007ly}, large pulsar kicks \cite{Kusenko:1999rt,Barkovich:2004ul,Fuller:2003vn,Loveridge:2004fr,Kishimoto:2011fk}, baryogenesis \cite{Akhmedov:1998ys,Asaka:2005fr}, core collapse supernova explosions \cite{Hidaka:2006yq, Fryer:2006rt,Hidaka:2007kx,Fuller:2009uq}, and other issues \cite{Munyaneza:2006ys}. (See a review on sterile neutrinos in Ref.~\cite{Kusenko:2009lr}.) The results of some accelerator-based experiments have been interpreted as suggesting active-sterile neutrino flavor mixing \cite{Athanassopoulos:1995vn,Athanassopoulos:1996yq,Athanassopoulos:1998kx,Athanassopoulos:1998fj,Aguilar:2001fj,Adamson:2009uq,Aguilar-Arevalo:2009qy,Aguilar-Arevalo:2010fk,Abe:2011uq,MiniB:2011kx,MiniB:2011lr}.

The seesaw mechanism \cite{Minkowski:1977zr,Yanagida:79,Gell-Mann:80,Glashow:1980,Mohapatra:1980mz} can explain small active neutrino masses by postulating order one Yukawa couplings to the Higgs and very large right-handed neutrino masses.  However, the split seesaw mechanism \cite{Kusenko:2010qy} can reconcile the small active neutrino masses and a sterile neutrino with a mass below the electroweak scale.  Ideas on what might be the rest masses of sterile neutrinos abound \cite{de-Gouvea:2005fk,de-Gouvea:2007lr}. 

For example, one class of models for generating the observed active neutrino masses \cite{Smirnov:2006uq} invokes heavy sterile neutrinos to induce a contribution to the lighter neutrino mass eigenvalues ({\it i.e.,} those most closely associated with the active neutrinos). Interestingly, in our calculations of radiation energy density we find that the ranges of sterile neutrino rest mass and vacuum mixing (with active species) which give the best agreement with observation sometimes fall in the \lq\lq sweet spot\rq\rq\ ranges which explain the observed neutrino masses in these models.

Active-sterile neutrino mixing results in effective sub-weak interaction coupling between sterile neutrinos and the rest of the universe \cite{Fuller:2009uq}. The upshot is that the sterile states may be thermally populated in the early universe, and these sterile neutrinos can decay to leptons and photons. The key issue is that sterile neutrinos which mix with active species can play havoc with astrophysical environments where neutrinos influence energetics and composition. A case in point is the early universe.  

There are at least three classic observational avenues to probe or constrain a generation of particles in the early universe that have decayed away. First, the observationally-inferred light element abundances, when confronted with standard BBN calculations which utilize the CMB anisotropy-inferred baryon-to-photon ratio, could reveal discordance with observationally-inferred primordial light element abundances. Such discordance could, in turn, lead to constraint. 

Second, the decay products of these particles may not completely thermalize, leading to extra radiation energy density not accounted for by photons and the three active neutrino flavors. For example, WMAP7 \cite{Komatsu:2011lr} reports $N_{\rm eff} = 4.34^{+0.86}_{-0.88}$, where the radiation energy density $\rho_{\rm rad}$ is related to the photon temperature $T$ through 
\begin{equation}
\rho_{\rm rad} = {\left[ 2+\frac{7}{4}{{\left( {{4}\over{11}} \right)}^{4/3}} N_{\rm eff} \right]} {{\pi^2}\over{30}} T^4.
\label{Neffdef}
\end{equation}
A recent analysis of CMB and large scale structure data argues that $N_{\rm eff} = 4.08^{+0.71}_{-0.68}$ \cite{Archidiacono:2011rt}. 
Note that $N_{\rm eff}$ is simply a parametrization of the radiation energy density. In principle this parameter might have nothing to do with neutrinos, though the choice of factors in its definition is tied to the standard cosmology scenario in which it does. For example, the factor of ${\left({{T_\nu}/{T}}\right)}^{4/3}=(4/11)^{4/3}$ in this definition comes from the standard cosmology case: Fermi-Dirac-shaped energy distribution functions for the ordinary relic neutrino backgrounds with \lq\lq temperature\rq\rq\ $T_\nu$ and degeneracy parameter zero. In the standard case $N_{\rm eff}$ should be close to $3$, matching the number of active neutrino flavors. (A careful calculation in the standard cosmology case, accounting for de-coupling induced deviations from thermal distribution functions for the neutrinos, but retaining the definition of $N_{\rm eff}$ in Eq.~\ref{Neffdef}, yields $N_{\rm eff} = 3.046$ \cite{Calabrese:2011lr}.) 

Third, an epoch of matter domination by these particles in the early universe could alter the growth or damping of structures on scales smaller than that of the causal horizon at this epoch. Of course, this is of little use as a means of constraining the particles considered here because the mass-scale of fluctuations so modified would be quite small, well below what can be probed via the CMB and current observations of small-scale, large-scale structure, {\it e.g.,} the Lyman alpha forest. 


Here we assess constraints on sterile neutrino rest mass and lifetime parameter space mostly from the second of these considerations. To do this we sought to answer a seemingly straightforward question: What would be the effects on the thermal history of the early universe stemming from heavy sterile neutrinos with number densities of order those of photons but which decay at or around the BBN epoch in the early universe? We targeted sterile neutrinos with rest masses ($100$-$500\,{\rm MeV}$) very large compared to the BBN energy scale ($\sim 1\,{\rm MeV}$), with vacuum mixing with active neutrinos sufficiently small that these species evade current laboratory neutrino mass bounds and drop out of thermal and chemical equilibrium at epochs with temperature $T> 1\,{\rm GeV}$, but sufficiently large that they would decay away during the BBN epoch. It is likely that the gross violence such a scenario would do to the time-temperature and scale factor-temperature relations and neutron-to-proton ratio during BBN would render the light element abundances sufficiently different from measured values so as to constrain this largely unexplored sector of neutrino physics. 

Even without light element BBN considerations, the work reported here demonstrates that a large range of the sterile neutrino rest mass and mixing parameter space, unconstrained by existing laboratory experiments, can nevertheless be ruled out via CMB limits on $N_{\rm eff}$. 

Curiously, however, we find that there are ranges of these sterile neutrino parameters where $N_{\rm eff}$ is consistent with observational constraints, but where the usual cosmic relic active neutrino background makes a negligible contribution to $N_{\rm eff}$ and, instead, the dominant contribution to this quantity arises from the active neutrino decay products of the sterile neutrinos. We leave to a later paper whether scenarios with {\it these} sterile neutrino parameters can evade BBN-based primordial light element abundance bounds.

Scenarios along these lines with decaying sterile neutrinos that influence the relic neutrino background open a fourth channel for constraint: laboratory measurement of a neutrino rest mass and the comparison of this with the CMB- and structure-derived cosmological bounds on neutrino mass. A robust signature of the sterile neutrino decay scenario outlined here would be a measured $N_{\rm eff} \neq 3$ and a measurement of neutrino mass in, {\it e.g.,} KATRIN or a neutrino-less double beta decay experiment, exceeding the cosmological bound on neutrino mass. 

In fact, it is better than this because we already know (from the square-root of the atmospheric neutrino mixing mass-squared splitting) that at least one neutrino mass eigenvalue exceeds $50\,{\rm meV}$ in the normal neutrino mass hierarchy, and two in the inverted hierarchy. As cosmological neutrino mass constraints push down toward $100\,{\rm meV}$ \cite{Abazajian:2011fk} the stage is set for a definitive test of the entropy generation/dilution scenarios discussed here.    


We identify roughly the regions of sterile neutrino rest mass and decay lifetime ({\i.e.,} coupling or mixing with active neutrinos) that produce matter dominated regimes that may last many Hubble times and where entropy production can be significant. On this latter point, the mechanism of out of equilibrium sterile neutrino decay-generated entropy production and its phasing with weak freeze-out and BBN is interesting in itself.


In what follows we address current bounds on sterile neutrino masses and vacuum flavor mixing in section II, sterile neutrino decoupling in section III, entropy production and associated dilution from out-of-equilibrium sterile neutrino decay in section IV,  $N_{\rm eff}$ modification resulting from these decay scenarios and constraints in section V, speculation on the significance of \lq\lq hidden\rq\rq\ matter-dominated epochs in section VI, and dilution effects on the cosmological determination of active neutrino masses in section VII. We give conclusions in section VIII.

\section{Constraints on Sterile Neutrino Rest Masses and Flavor Mixing}

\subsection{Active neutrino mass limits}

Sterile neutrinos which mix in vacuum with active neutrinos are not truly sterile because they have an effective interaction with strength $\sim \sin^2\theta\ G_{\rm F}^2$, where $\theta$ is the effective vacuum mixing angle and the Fermi constant squared $G_{\rm F}^2$ gives the strength of the ordinary weak interaction. This effective coupling, though perhaps very small, nonetheless provides an avenue for these particles to be created in the early universe, and constrained by direct laboratory experiments and observations.

First consider the constraints. Laboratory neutrino mass constraints \cite{Otten:2008mz,Drexlin:2008zr}, particularly those based on the tritium beta decay endpoint, which are good whether neutrinos are Majorana or Dirac in character, also constrain the extent to which very massive sterile neutrinos can mix with, for example, the electron neutrino in vacuum. The effective $\nu_e$ \lq\lq mass\rq\rq\ in vacuum is
\begin{equation}
m_{\nu_e} = \sum_i{{\vert U_{e\,i}\vert}^2 \, m_i} \approx {\vert U_{e\,4}\vert}^2\, m_4 \equiv m_s\, \sin^2\theta,
\label{lim}
\end{equation}
where $U_{e\, i}$ are the mass basis to flavor basis unitary transformation matrix elements, $m_i$ are the neutrino mass eigenvalues ($i=1,2,3,4$) and we minimally assume a fourth of these, with the sterile neutrino mass roughly $m_s=m_4$. Here we consider the limit where $m_4 \gg m_{1,2,3}$, so that the effective vacuum active-sterile mixing angle is defined through $\sin^2\theta = {\vert U_{e\,4}\vert}^2$. If $m_{\nu_e}^{\rm lim}$ is the tritium endpoint experimental upper limit on neutrino mass, then we must have $\sin^2\theta < m_{\nu_e}^{\rm lim}/m_s$.

Sterile neutrino masses less than the pion rest masses ($m_{\pi^0} = 135\,{\rm MeV}$, $m_{\pi^{\pm}} =139.57\,{\rm MeV}$) will decay primarily through three neutrino decay $\nu_s\rightarrow \nu_\alpha + \nu_\beta+\bar\nu_\beta$, while those above this mass-scale will decay via $\nu_s \rightarrow \pi^0 +\nu_\alpha$, with $\alpha,\beta=e,\mu,\tau$, or via various charged pion decay modes that will be discussed below. The three-neutrino decay channel, for example, leads to $\nu_s$ lifetime \cite{Pal:1982qy,Abazajian:2001lr,Abazajian:2001fk}
\begin{equation}
\tau_{3\nu} \approx 2.88\times{10}^4\,{\rm s} \,{\left({{\rm MeV }\over{m_s }}\right)^5}\,{{1}\over{\sin^2\theta }},
\label{3nudecay}
\end{equation}
while the $\pi^0$ decay channel, for example, produces an inverse decay rate which very roughly is  
\begin{equation}
\tau_{\pi^0} \approx 5.76\times{10}^{-9}\,{\rm s} \,{{1}\over{\sin^2\theta }}\cdot{{1}\over{x\,{\left( x^2-1\right)} }},
\label{nupidecay}
\end{equation}
where $x\equiv m_s/m_{\pi^0}$. The other pion decay channels will be discussed below, but this one is adequate to illustrate rough laboratory $m_s$ constraints.

Neutrino mass upper limits from any laboratory source translate into lower bounds on the lifetimes of sterile neutrinos, $\tau$: for the three neutrino decay mode,
\begin{equation}
\tau_{3\nu}\, >\, 1.44\times{10}^{10}\,{\rm s}\,{\left( {{ 2\,{\rm eV}}\over{m_{\nu_e}^{\rm lim} }} \right)}{\left( {{\rm MeV}\over{m_s }} \right)}^4;
\label{tau3nu}
\end{equation}
 while for the $\pi^0$ decay mode
 \begin{equation}
\tau_{\pi^0}\, >\, 0.389\,{\rm s}\,{\left( {{2\,{\rm eV}}\over{ m_{\nu_e}^{\rm lim} }} \right)}\cdot{{1}\over{ x^2-1}}.
\label{tau3nu}
\end{equation}
Neither of these constraints affects the sterile neutrino mass and lifetime parameter space of most interest in this paper. Accelerator-based laboratory constraints may impact the lowest mass and highest mixing angle sterile neutrinos considered here \cite{Kusenko:2005qy,Bergsma:1983uq,Bernardi:1986fj,Bernardi:1988kx,Baranov:1993yq,Nedelec:2001vn}. Future tritium endpoint experiments like KATRIN may take the $\nu_e$ mass limit down by an order of magnitude, to $m_{\nu_e}^{\rm lim} \approx 0.2\,{\rm eV}$, with appropriately more stringent lower limits on sterile neutrino lifetimes. Double beta decay experiments may provide far better limits, but these apply only to Majorana neutrinos. 

Cosmological constraints (based on large scale structure and the CMB) on the sum of the light neutrino masses \cite{Abazajian:2011fk,Kaplinghat:2003bh,Hannestad:2004ly,de-Bernardis:2009fr,Hannestad:2010ys,Archidiacono:2010ve} will come into our considerations in a manner very different from the usual arguments in standard cosmology. The active neutrinos decouple early ($\sim 1\,{\rm s}$) and the usual cosmological constraints are figured at photon decoupling ($t\sim {10}^{13}\,{\rm s}$) or later, and at these epochs the sum of the light neutrino masses $M_\nu$ will have little contribution from heavy sterile states, as these components will have decayed away. At time $t$ we would have 
\begin{equation}
M_\nu \approx \sum_{\alpha=e,\mu,\tau}\,\sum_{i=1,2,3}\,{{\vert U_{\alpha\, i}\vert}^2 \, m_i}+{\vert U_{e\,4}\vert}^2\, e^{-t/\tau}\, m_4,
\label{decaylim}
\end{equation}
if the background neutrinos were in flavor states satisfying all the usual cosmological assumptions and if mass state $4$ mixes only with $\nu_e$ in vacuum, with obvious generalizations for more complicated active-sterile mixing matrix realizations.

There are stringent cosmological constraints on long-lived extra sterile neutrino states, see {\it e.g.,} Ref.~\cite{Giusarma:2011gf}. For the most part, these cosmological constraints are predicated on the {\it assumptions} of relic active neutrino backgrounds with black body, thermal-shaped, Fermi-Dirac energy spectrum characterized by a temperature some $40\%$ lower than the photon temperature, and zero chemical potentials. As we will show below, the sterile neutrino-decay scenarios considered here produce a very different relic active neutrino background.

\subsection{Neutrino Mass Models}

In some models for the origin of neutrino rest mass the ultimate mass eigenvalue is a sum of an intrinsic piece and a contribution induced by the presence of a massive (mostly) sterile neutrino component as described in Ref.~\cite{Smirnov:2006uq}. This induced mass contribution is $\sim m_s\,\theta^2$, so that there are three broad categories we can consider. 

First is the regime where this product is much less than the known active neutrino mass scale, {\it i.e.,} $m_s\,\theta^2 \ll 0.05\,{\rm eV}$, or where $\sin^22\theta < {10}^{-10}/\left( m_s/100\,{\rm MeV}\right)$. These neutrino mass models give effectively no constraint in this regime. Many if not most of the sterile neutrino mass and mixing/lifetime parameters we consider below fall in this regime, especially for $m_s > 150\,{\rm MeV}$

Second is the regime where $0.05\,{\rm eV} < m_s \theta^2 < 0.5\,{\rm eV}$, or $\sin^22\theta \sim \left( {10}^{-10} \-- {10}^{-9} \right)/\left( m_s/100\,{\rm MeV}\right)$. This is right in the range to give the requisite induced mass contribution to the active neutrinos. As will will see, this range actually overlaps with some of the ranges of sterile neutrino rest mass and mixing which give the best agreement with observational constraints on $N_{\rm eff}$.

Finally, a third regime is where $m_s\,\theta^2 \gg 0.5\,{\rm eV}$, corresponding to  $\sin^22\theta > {10}^{-9}/\left( m_s/100\,{\rm MeV}\right)$. Sterile neutrino rest mass and lifetime/mixing parameters falling in this range would be disfavored in these induced neutrino mass models.

\subsection{Considerations from Colliders}

Since the sterile neutrino has non-zero mixing with the active neutrino flavors, any interaction that produces an active neutrino can instead produce a sterile neutrino, with a cross section reduced by the effective vacuum active-sterile mixing, $\sin^2 \theta$.  If a sterile neutrino is created in a collider and then subsequently decays in a detector, a signature for the sterile neutrino may be inferred \cite{Shoemaker:2010lr}.  For example, the decay of sterile neutrinos with mass larger than $m_{\pi^\pm}$ will produce a co-linear pion-electron pair.

The rate of sterile neutrino decays in the detector can be estimated from the overall sterile neutrino production rate by taking the product of the collider luminosity, $L$, the weak cross section, $\sigma \sim G_F^2 E^2$, and the vacuum mixing angle suppression factor, $\sin^2 \theta$.  The detection rate would be the production rate, suppressed by the ratio of the crossing time of the neutrino through the detector to the laboratory-frame lifetime of the sterile neutrino (product of Lorentz factor $\gamma$ and $\tau$)
\begin{align}
\frac{R}{\gamma \tau}  \sim & 10^{-8} \left( \frac{R}{30~{\rm m}} \right) \left( \frac{E_{\nu_s}}{30~{\rm GeV}} \right)^{-1} \nonumber \\
 & \qquad \times \left( \frac{m_s}{200~{\rm MeV}} \right) \left( \frac{\tau}{100~{\rm s}} \right)^{-1} .
\end{align}
Detection or constraints along these lines obtained from colliders would be difficult, but not impossible.

\section{Sterile Neutrino Decoupling in the Early Universe}

The sterile neutrinos discussed above can have effective interactions large enough to keep them in thermal and chemical equilibrium for epochs in the early universe with temperatures sufficiently high. For the rest mass ranges we consider here these species will have relativistic kinematics when they decouple. Therefore, subsequently, they will have a momentum space distribution function that is a relativistic Fermi-Dirac black body, for which we will take the chemical potential to be zero. Of course, as the universe expands, coupled particles will annihilate or decrease in number and the entropy that these particles carry will be transferred to the still-coupled particles and not to the decoupled sterile neutrinos. The net result will be that at the epoch of standard weak decoupling, temperature $T\sim 1\,{\rm MeV}$, the sterile neutrinos will have a \lq\lq temperature\rq\rq\ lower than that of the active neutrinos and the plasma.

At early epochs, when the sterile neutrinos are coupled and possess relativistic kinematics, their scattering rate will be $\sim G_{\rm F}^2\,T^5\,\sin^22\theta$. Thermal equilibrium between the sterile neutrinos and the plasma will obtain where this rate is larger than the local Hubble expansion rate. In radiation dominated conditions the expansion rate is $H ={\left( 8\pi^3/90 \right)}^{1/2}\ g^{1/2}\ T^2/m_{\rm pl}$, where $g$ is the statistical weight in relativistic particles and $m_{\rm pl}$ is the Planck mass. Comparing these rates, the temperature where sterile neutrinos decouple is roughly
\begin{eqnarray}
T_{\rm dec} & \approx & {\left( {{ 8 \pi^3}\over{90 }} \right)}^{1/6}\,{{ g^{1/6}}\over{{\left( G_{\rm F}^2\, m_{\rm pl} \right)}^{1/3}\, \sin^{2/3}2\theta }}\\
&\approx &2\,{\rm GeV}\,{\left( {{ g}\over{61.75 }} \right)}^{1/6}\,{\left({{{10}^{-9}}\over{\sin^{2}2\theta}}\right)}^{1/3}.
\label{dec}
\end{eqnarray}
For example, a sterile neutrino with rest mass $m_s = 200\,{\rm MeV}$ and $\sin^22\theta={10}^{-10}$ would have lifetime $\tau \approx 43\,{\rm s}$, and between its decoupling at $T\sim4\,{\rm GeV}$, and weak (active neutrino) decoupling at $T\approx 3\,{\rm MeV}$, a tenth of a second or so would have elapsed, and few sterile neutrinos would have decayed. Instead, though they would then have largely nonrelativistic kinematics, their energy spectrum would have a relativistic Fermi-Dirac form, albeit with a temperature lower than that for the plasma and active neutrinos.

Where particle decay is negligible the entropy in a co-moving volume will be conserved. If the entropy is carried by relativistic particles the ratio of the sterile neutrino temperature to the active neutrino and plasma temperatures at weak decoupling (wd) will be
\begin{equation}
{{T_{\nu_s}}\over{T}}\Big\vert_{\rm wd} ={\left( {{ g_{\rm wd}}\over{ g_{\nu_s{\rm dec}} }} \right)}^{1/3}\approx{\left( {{10.75}\over{61.75 } } \right)}^{1/3} = {{1}\over{1.79}},
\label{ratioTs}
\end{equation}
where $g_{\nu_s{\rm dec}}$ and $g_{\rm wd}$ are the statistical weight in relativistic particles at the epochs of sterile neutrino and weak decoupling, respectively. The former quantity is figured to be roughly $g_{\nu_s{\rm dec}} \approx g_{\gamma e \mu\nu}+g_{\rm qg}$, where the statistical weight in photons, neutrinos, $e^\pm$, and $\mu^\pm$ is $g_{\gamma e \mu\nu} = 14.25$, while that in quark and gluon degrees of freedom is $g_{\rm qg} = 16+10.5\,N_{\rm f}=47.5$, where we leave out tau leptons in the former and take three relativistic quark flavors, $N_{\rm f} =3$ ($u,d,s$), in the latter weight.
By far the biggest source of dilution in $T_{\nu_s}/T$ stems from the loss of quark and gluon degrees of freedom at the QCD epoch at $T\approx 170\,{\rm MeV}$.

Since the ordinary thermal background neutrinos have roughly the same temperature as photons at the beginning of weak decoupling, Eq.~\ref{ratioTs} implies that the ratio of sterile neutrino to active background neutrino temperature is 
\begin{equation}
G \equiv {{T_{\nu_s}}\over{T_\nu}} \approx {{T_{\nu_s}}\over{T}}\Big\vert_{\rm wd}. 
\label{G}
\end{equation}
Since the decoupled sterile and ordinary thermal background neutrino energies redshift with scale factor in the same way, $G$ will be a co-moving invariant, {\it i.e.,} fixed.

Despite dilution from epochs of particle annihilation, at weak decoupling the number density of sterile neutrinos, assuming negligible numbers of decays, will still be a significant fraction of the photon number density $n_\gamma$,
\begin{equation}
n_{\nu_s+\bar\nu_s}\Big\vert_{\rm wd} \approx {{3}\over{2}} {{\zeta\left(3\right)}\over{\pi^2}} T_{\nu_s}^3 \sim 0.1\, n_\gamma,
\label{numbdens}
\end{equation}
where the Riemann Zeta function is $\zeta(3) \approx 1.20206$ and we include both left- and right-handed sterile neutrinos. As time goes on these sterile neutrinos will decay away. The actual proper number density of these particles at time $t$ is
\begin{equation}
n_{\nu_s+\bar\nu_s} ={{3}\over{2}} {{\zeta\left(3\right)}\over{\pi^2}} T_{\nu_s}^3\,e^{-t/\tau}.
\label{numbdecay}
\end{equation}
Of course, there will be further dilution of $T_{\nu_s}/T$ as $e^\pm$ pairs annihilate and entropy is added to the photon/baryon/electron plasma through sterile neutrino decay.

\section{Decay-Induced Entropy Generation and Dilution}

Heavy sterile neutrinos can decay out of equilibrium in the early universe and thereby generate entropy and cause what is sometimes termed {\it dilution}. These thermodynamic and cosmological issues were discussed in the general case for any unstable particles in Ref.~\cite{Scherrer:1985zr}. Here we examine dilution for the specific case of heavy sterile neutrinos decaying around the BBN epoch, but there are new twists in this case. For example, there are sterile neutrino decay channels in which ordinary active neutrinos are produced. These decay-produced active neutrinos may or may not thermalize. If they thermalize they add to dilution. If they do not they add to the sea of decoupled relativistic particles. Taken together, dilution and the addition of decoupled decay neutrinos affect $N_{\rm eff}$ in competing ways.

\subsection{Sterile Neutrino Decay Processes}

Massive sterile neutrinos can decay through flavor mixing with active species as outlined above.  If the sterile neutrino mass is less than the $\pi^0$ mass, then these particles will decay into three active neutrinos, with a smaller branch into a neutrino and a photon, or into a neutrino and an $e^\pm$-pair. For sterile neutrinos with rest masses above the $\pi^0$ or $\pi^\pm$ rest mass scales other, more complicated decay channels become possible.  These are discussed below and in Ref.s~\cite{Abazajian:2001lr,Abazajian:2001fk} and \cite{Pal:1982qy,Barger:1995lr}.

For each decay mechanism, a fraction $f_{\rm em}$ of the total sterile neutrino decay energy will be comprised of electromagnetic decay products (photons, electrons and positrons). This fraction of the decay energy will be added to the plasma through inelastic scattering of these products on particles in the photon-electron-baryon plasma.  This is because the scattering timescale for these particles is much shorter than the dynamical timescale.  

On the other hand, the energy contained in the active neutrino decay products will remain in the neutrino sector, unless the individual neutrino energies are large enough that their scattering timescales rival the dynamical timescale. In the calculations presented here we compare the decay-produced neutrino scattering rate to the local Hubble expansion rate and estimate the neutrino energy where these rates are equal,
\begin{equation}
E_\nu^{\rm retherm} = {\left({{8}\over{3}} \pi\ \rho  \right)}^{1/2}\, {\left[  G_{\rm F}^2\, m_{\rm pl}\, T^4   \right]}^{-1}
\label{Edec},
\end{equation}
where $\rho$ is the total mass-energy density, $G_{\rm F}$ is the Fermi constant, and $m_{\rm pl}$ is the Planck mass.

If a decay neutrino has an energy larger than $E_\nu^{\rm retherm}$ we assume that it will down-scatter until it has energy $E_\nu^{\rm retherm}$. The difference between its initial energy and $E_\nu^{\rm retherm}$ will be deposited in the plasma. The fraction of the sterile neutrino rest mass deposited in the plasma via the scattering of decay neutrinos is $f_\nu$, with $f=f_{\rm em}+f_\nu$. Decay neutrinos with energies smaller than $E_\nu^{\rm retherm}$ are taken to be decoupled. Neutrinos which are decoupled do not contribute to the thermalization fraction $f$, but they do contribute to relativistic energy density, {\it i.e.,} to $N_{\rm eff}$. 

There are seven decay processes included in the calculations presented here. These are: 

\subsubsection{$\nu_s \rightarrow 3 \nu$}

The decay rate for this process is
\begin{eqnarray}
\Gamma_{3\nu} & = &  {{G_{\rm F}^2}\over{192\,\pi^3}}\cdot m_s^5\cdot\sin^2\theta\\
                             & \approx & 3.47\times{10}^{-5}\,{\rm s}^{-1}\cdot {\left({{m_s}\over{\rm MeV}}\right)}^5\cdot \sin^2\theta.
\label{3nudecay}
\end{eqnarray}
We make the assumption here that the heavy state $\nu_4$ decays through a coupling with one active species with effective vacuum mixing angle $\theta$. This is tantamount to a particular mass basis/flavor basis transformation, with implications for degeneracy in the decay rate. 
The final state active decay neutrinos produced in this process usually are decoupled and so deposit no energy in the photon-electron-baryon plasma, {\it i.e.,} $f=0$. The exception is for the most massive sterile neutrinos considered here, and only then very early in the expansion.

\subsubsection{$\nu_s \rightarrow \nu+\gamma$}

Here we follow Ref.s~\cite{Abazajian:2001lr} and \cite{Abazajian:2001fk} and take the rate for this process as
\begin{eqnarray}
\Gamma_{\nu_s \gamma} & = & \alpha\cdot {{G_{\rm F}^2  }\over{64\, \pi^4 }}\cdot m_4^5 \cdot {\left[ \sum_\beta{ U_{1 \beta}\,U_{4 \beta}\, F\left( r_\beta\right)}  \right]}^2\\
& \approx &  {{9 G_{\rm F}^2}\over{512 \,\pi^4}}\cdot \alpha \cdot m_s^5\cdot\sin^2\theta\\
                             & \approx & 2.72\times{10}^{-7}\,{\rm s}^{-1}\cdot {\left({{m_s}\over{\rm MeV}}\right)}^5\cdot \sin^2\theta,
\label{nugamdecay}
\end{eqnarray}
where $\alpha$ is the fine structure constant and we assume a particular vacuum flavor mixing structure, and again use $m_s\equiv m_4$. This electromagnetic decay process is not GIM-suppressed because it involves sterile neutrinos with no corresponding charged lepton. In these expressions, $F\left( r_\beta\right)\approx -\frac{3}{2}+\frac{3}{4} r_\beta$, $r_\beta \approx {\left( m_\beta/M_{\rm W}  \right)}^2$, the ratio of charged lepton mass to the ${\rm W}$ mass, all squared, and the sum is over flavor $\beta$. Because the sterile neutrinos considered here have non-relativistic kinematics at relevant epochs, while the decay products are relativistic, the final state neutrino and photon each have energy $m_s/2$. This implies that $f\approx 1/2$ for this process. 

\subsubsection{$\nu_s \rightarrow \nu+e^++e^-$}

When the sterile neutrino rest mass is much larger than twice the electron rest mass, then the rate for this decay process is just $1/3$ of the rate for the 3-neutrino decay process,
\begin{equation}
\Gamma_{\nu e^+ e^-} = {{1}\over{3}}\,\Gamma_{3\nu}.
\label{nuepmdecay}
\end{equation}
The ratio of $1/3$ between the rates of two processes comes from the difference in the decay of the virtual $Z^0$ in these cases: the decay is into three flavors of neutrino pairs in 3-neutrino decay; and into a single $e^\pm$-pair in the $\nu_s\rightarrow \nu +e^-+e^+$ channel. We will have roughly $f\approx 2/3$ for this process when $m_s$ is large enough.

\subsubsection{$\nu_s \rightarrow \nu+\mu^++\mu^-$}

Similarly, when the sterile neutrino rest mass is much larger than twice the muon rest mass ($m_s \gg 2\ m_\mu$) the rate for this decay process is the same as for the last process,
\begin{equation}
\Gamma_{\nu \mu^+ \mu^-} = \Gamma_{\nu e^+ e^-}.
\label{nuepmdecay}
\end{equation}
Obviously this decay channel has a significant sterile neutrino rest mass threshold, $m_s > 2 m_\mu \approx 211.32\,{\rm MeV}$. Moreover, the plasma energy deposition in this decay channel is different because the muons decay. For example, $\mu^-\rightarrow e^-+\nu_\mu+\bar\nu_e$ produces an electron and two neutrinos. The electron or positron thermalizes while the neutrinos, each of which has average energy $E_{\nu\mu}=34.33\,{\rm MeV}$, usually remain decoupled and so do not.

\subsubsection{$\nu_s \rightarrow \pi^0 +\nu$}

Sterile neutrinos with masses larger than the $\pi^0$ mass ($m_s > m_{\pi^0} \approx 135~{\rm MeV}$) can decay into a $\pi^0$ and an active neutrino.  The $\pi^0$ then decays into two photons. This decay channel has mean lifetime $8.4 \times 10^{-17}~{\rm s}$, which is much shorter than both the scattering timescale and dynamical timescale in the early universe, so we can regard these decays as being instantaneous.  The energy of the active neutrino is
\begin{equation}
E_\nu^{\pi^0} = \frac{m_s^2 - m_{\pi^0}^2}{2 m_s} .
\end{equation}
The remaining available decay energy, $m_s - E_\nu^{\pi^0}$, is thermalized into the plasma as the $\pi^0$ decays into photons. The decay rate for this channel is
\begin{eqnarray}
\Gamma_{\pi^0 \nu} 
& = &  {{G_{\rm F}^2 \, f_\pi^2}\over{16\,\pi}}\cdot{m_s \left[ m_s^2-m_{\pi^0}^2 \right]}\cdot\sin^2\theta\\
& \approx & 1.735\times{10}^{8}\,{\rm s}^{-1}\, x \,{\left[ x^2-1  \right]} \cdot \sin^2\theta,
\label{pi0nudecay}
\end{eqnarray}
with $x\equiv m_s/m_{\pi^0}$ and here we take $f_\pi =131\,{\rm MeV}$. 


\begin{figure}
\includegraphics[width=3.5in]{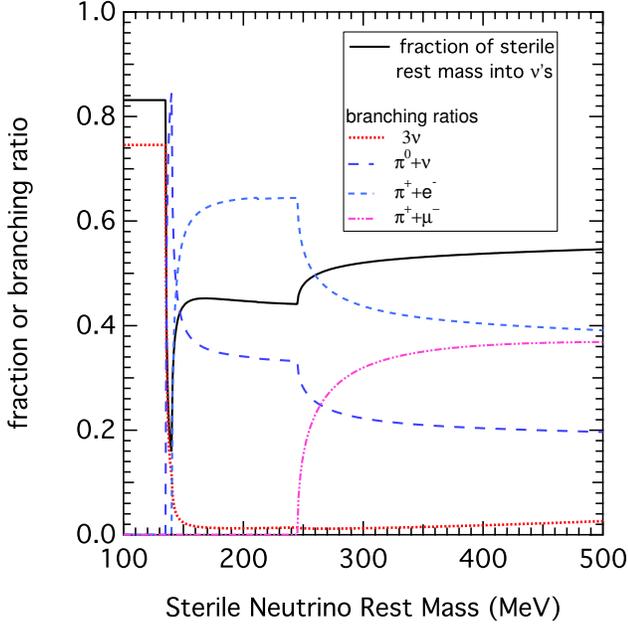}
\caption{Branching ratios for selected decay processes and the branching ratio-weighted fraction of sterile neutrino rest mass (neglecting thermalization) deposited in the decoupled active neutrino seas are shown as functions of sterile neutrino rest mass in MeV. }
\label{branchratio}
\end{figure}

\subsubsection{$\nu_s \rightarrow \pi^\pm +e^\mp$}

Sterile neutrinos with masses larger than the sum of the charged pion and electron masses ($m_s > m_{\pi^\pm} + m_e \approx 140.08~{\rm MeV}$) can decay into a $\pi^\pm$ and an $e^\mp$.  The electron (or positron) has energy
\begin{equation}
E_e = \frac{m_s^2 - m_{\pi^+}^2+m_e^2}{2 m_s} ,
\end{equation}
which is quickly thermalized into the plasma.  The charged pion then decays, {\it e.g.,} $\pi^+ \rightarrow \mu^+ +\nu_{\mu}$ (or its {\it CP}-conjugated counterpart), with a mean lifetime of $2.6 \times 10^{-8}\,{\rm s}$, which is much shorter than the scattering and dynamical timescales, so we can also regard these decays as effectively instantaneous. The muon then decays producing two neutrinos as described above. Altogether three neutrinos are produced in this decay. Charged pion decay at rest produces a mono-energetic $\nu_\mu$ 
(or $\bar\nu_\mu$) with energy $E_{\nu\pi} =  29.79\,{\rm MeV}$. The average energy added to the decoupled neutrino sea per decay in this channel is taken to be $\langle E_{\rm dec}\rangle =E_{\nu\pi}+2 E_{\nu\mu}+({5/6}) (E_{\pi {\rm KE}})$, where the kinetic energy of the pion is  $E_{\pi {\rm KE}}=m_s- E_e -m_{\pi^\pm}$.

\begin{widetext}

The decay rate for this channel is
\begin{eqnarray}
\Gamma_{\pi^+ e^-} & = &  {{G_{\rm F}^2 \, f_\pi^2}\over{16\,\pi}}\cdot{m_s \left[\left(m_s^2-{\left( m_{\pi^\pm}+m_e \right)}^2  \right)\left( m_s^2-{\left( m_{\pi^\pm}-m_e \right)}^2 \right)  \right]^{1/2}}\cdot\sin^2\theta\\
                             & \approx & 1.917\times{10}^8\,{\rm s}^{-1}\ {\left( {{ m_s}\over{\rm MeV }} \right)^3\cdot \left[\left(1-{\left( x+y \right)}^2  \right)\left( 1-{\left( x-y \right)}^2 \right)  \right]^{1/2}} \cdot \sin^2\theta,
\label{pi-e-decay}
\end{eqnarray}
with $x+y\equiv \left( m_{\pi^\pm}+m_e \right)/m_s$ and $x-y\equiv \left( m_{\pi^\pm}-m_e \right)/m_s$ and we again take $f_\pi =131\,{\rm MeV}$. Since there are two channels for this decay, {\it i.e.,} into $\pi^+$ or $\pi^-$, we take the total decay rate as $\Gamma_{\pi e}=2 \Gamma_{\pi^+ e^-}$.

\subsubsection{$\nu_s \rightarrow \pi^\pm+ \mu^\mp$}

The kinematics of this decay are similar to those of $\nu_s \rightarrow \pi^\pm+e^\mp$, but with the obvious difference in sterile neutrino mass threshold, which is now $m_s > m_{\pi^\pm} + m_\mu \approx 245.23{\rm MeV}$. The decay rate in this channel is
\begin{eqnarray}
\Gamma_{\pi^+ \mu^-} & = &  {{G_{\rm F}^2 \, f_\pi^2}\over{16\,\pi}}\cdot{m_s \left[\left(m_s^2-{\left( m_{\pi^\pm}+m_\mu \right)}^2  \right)\left( m_s^2-{\left( m_{\pi^\pm}-m_\mu \right)}^2 \right)  \right]^{1/2}}\cdot\sin^2\theta\\
                             & \approx & 1.917\times{10}^8\,{\rm s}^{-1}\ {\left( {{ m_s}\over{\rm MeV }} \right)^3\cdot \left[\left(1-{\left( x+y \right)}^2  \right)\left( 1-{\left( x-y \right)}^2 \right)  \right]^{1/2}} \cdot \sin^2\theta,
\label{pi-mu-decay}
\end{eqnarray}
with $x+y\equiv \left( m_{\pi^\pm}+m_\mu \right)/m_s$ and $x-y\equiv \left( m_{\pi^\pm}-m_\mu \right)/m_s$ and we again take $f_\pi =131\,{\rm MeV}$. Since there are two channels for this decay, {\it i.e.,} into $\pi^+$ or $\pi^-$, we take the total decay rate as $\Gamma_{\pi \mu}=2 \Gamma_{\pi^+ \mu^-}$.
\end{widetext}

The muon (or anti-muon) produced promptly in this decay has energy
\begin{equation}
E_\mu = \frac{m_s^2 - m_{\pi^+}^2+m_\mu^2}{2 m_s}.
\end{equation}
The charged pion co-produced with this muon has total energy $E_\pi=m_s-E_\mu$ and decays producing another muon, {\it e.g.,} $\pi^+ \rightarrow \mu^+ +\nu_{\mu}$ (or its {\it CP}-conjugated counterpart). Subsequently the two muons decay, each producing two neutrinos as described above. Altogether five neutrinos are produced in this decay sequence. The average energy added to the decoupled neutrino sea per decay in this channel is taken to be $\langle E_{\rm dec}\rangle =E_{\nu\pi}+4 E_{\nu\mu}+({5/6}) (E_{\pi {\rm KE}})+ ({2/3}) E_{\mu {\rm KE}}$, where the kinetic energy of the pion is  $E_{\pi {\rm KE}}=E_{\pi}-m_{\pi^+}$, while the kinetic energy of the prompt muon is $E_{\mu {\rm KE}}=E_\mu-m_\mu$.

A caveat: the list of sterile neutrino decay processes presented here is not definitive nor is it exhaustive. For example, the sterile-active vacuum flavor mixing scheme remains a complete guess, and this guess affects degeneracies in the rate calculations. (Do all active neutrino couple to the sterile in the {\it same} way, so that appropriate decay rates are multiplied by three?) Additionally, for sterile neutrino masses $m_s > 500\,{\rm MeV}$, five-neutrino decay becomes dominant \cite{Abazajian:2001lr} and even three-neutrino decay, with its five powers of $m_s$ dependence, will eventually overtake the pion decay channels with their three-powers dependence. Nevertheless, for the sterile rest mass ranges we consider here, where $N_{\rm eff}$ and dilution effects are most interesting, the seven decay channels we employ suffice to rough out the key behavior. 

Figure~\ref{branchratio} shows the branching ratios and the fraction of sterile neutrino energy deposited in the decoupled active neutrino sea all as functions of sterile neutrino rest mass. This figure does not show the (usually small) fraction of decay active neutrino energy which is thermalized. The sterile neutrino mass thresholds produce the obvious kings and ledges visible in this figure. For sterile neutrinos with masses below the $\pi^0$ threshold, three-neutrino decay dominates and only roughly $f\approx 17\%$ of the rest mass ends up thermalized in the plasma, and this principally through $\nu_s \rightarrow \nu+e^-+e^+$ and $\nu_s\rightarrow \nu+\gamma$. Above the $\pi^0$ and $\pi^\pm+e^\mp$ thresholds it is a different story, with $f\sim 60\%$ until the sterile neutrino mass exceeds the $\pi^\pm+\mu^\mp$ threshold where, because this decay channel produces {\it five} final state active neutrinos, $f$ falls to less than $50\%$. Of course, this also means that the majority of energy deposited in the decoupled active neutrino sea comes from this process when $m_s$ is high enough.  

\begin{figure}
\includegraphics[width=3.7in]{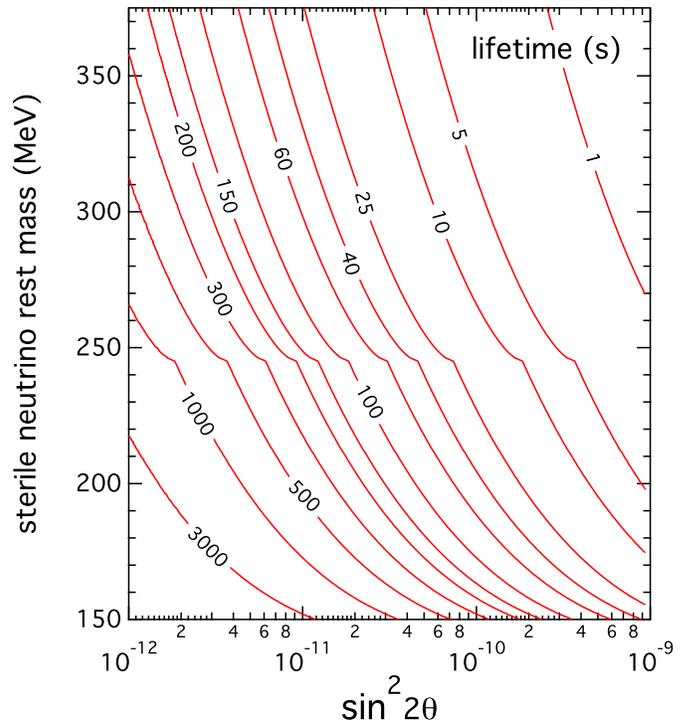}
\caption{Contours of sterile neutrino lifetime $\tau$ in seconds as functions of sterile neutrino rest mass $m_s$ and $\sin^22\theta$.}
\label{tau}
\end{figure}

\begin{figure}
\includegraphics[width=3.7in]{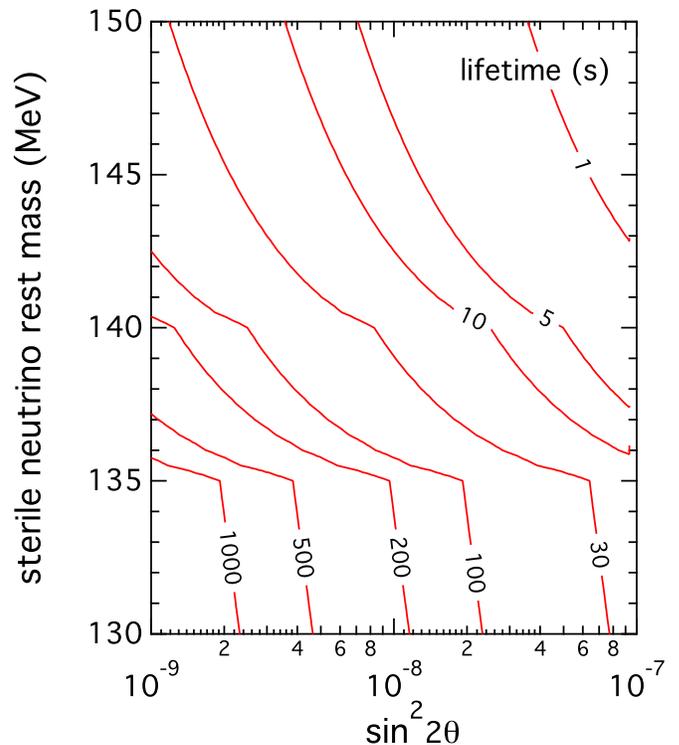}
\caption{Same as Fig.~\ref{tau}, but for lower ranges of sterile neutrino rest mass and larger vacuum mixing angles.}
\label{tau-L}
\end{figure}

Rates must be summed for all allowed decay channels to produce a total decay rate, {\it i.e.,} an inverse lifetime $1/\tau$. The relation between the overall sterile neutrino decay lifetime $\tau$ and the rest mass $m_s$ and $\sin^22\theta$ is shown in Fig.~\ref{tau} and Fig.~\ref{tau-L}. (In these figures we plot $\sin^22\theta$ instead of $\sin^2\theta$ for ease of comparison to several extant sterile neutrino constraint works.)

\subsection{Entropy Generation}

The relation between the entropy-per-baryon $S$, in units of Boltzmann's constant $k_{\rm b}$, and the baryon-to-photon ratio, $\eta\equiv n_b/n_\gamma$, is
\begin{eqnarray}
s & = & {\left( {{ \pi^4 }\over{ 45\,\zeta\left( 3\right)}} \right)}\,{{g_s}\over{\eta}}\\
& \approx & {\left( 5.895\times{10}^9\right)}\, {\left( {{g_s}\over{2}} \right)} {\left( {{ 6.11\times{10}^{-10}}\over{\eta }} \right)},
\label{entropy1}
\end{eqnarray}
where $g_s$ is the statistical weight in relativistic particles carrying the entropy, so that $S={\cal{S}}/n_b=(2\pi^2/45) g_s T^3/n_{\rm b}$, where $n_{\rm b}$ and $n_\gamma$ are the proper number densities of baryons and photons, respectively. In a standard universe at temperatures low enough that baryon number is conserved and with no particle decay, nuclear reactions, shocks, etc., $s$ would be constant. Note, however, that even in this limit the relation between $\eta$ and $s$ is not constant because $g_s$ changes with time/temperature. The Wagoner-Fowler-Hoyle $h$ parameter is defined as $h\equiv n_b/(N_{\rm A} T_9^3)$, where $N_{\rm A}$ is Avogadro's number and $T_9\equiv T/{10}^9\,{\rm K}$. With this definition the entropy-per-baryon is $s\approx 1.213\times{10}^5 (g_s/2)/h$ and $h=(3.368\times{10}^4\,{\rm g}\,{\rm cm}^{-3})\,\eta$, and the baryon rest mass density is $\rho_b [{\rm g}\,{\rm cm}^{-3}] = T_9^3\,h$. 

Decay of decoupled, out of equilibrium sterile neutrinos will result in entropy being added to the plasma. The rate at which entropy-per-baryon is added is
\begin{equation}
{{d s}\over{d t}} = {{m_s}\over{T}}\cdot f\cdot x_0\cdot e^{-\left(t-t_0\right)/\tau}\cdot{{1}\over{\tau}},
\label{sdot}
\end{equation}
where $f$ is the fraction of the sterile neutrino decay energy (the rest mass $m_s$) which thermalizes in the plasma. In Eq.~(\ref{sdot}) the number of sterile neutrinos ($\nu_s+\bar\nu_s$) per baryon at time $t_0$ is $x_0 = (3/4)(T_{\nu_s}/T)_0^3/\eta_0$, where the zero subscript means these quantities are evaluated at $t_0$, the age of the universe at which the calculation is started. In the calculations to follow we take $t_0$ sufficiently early ({\it e.g.}, well above the weak decoupling temperature) that prior decay can be neglected, in which case $x_0 \approx (3/4)(1/1.79)^3/\eta_0$. 

\begin{figure}
\includegraphics[width=3.5in]{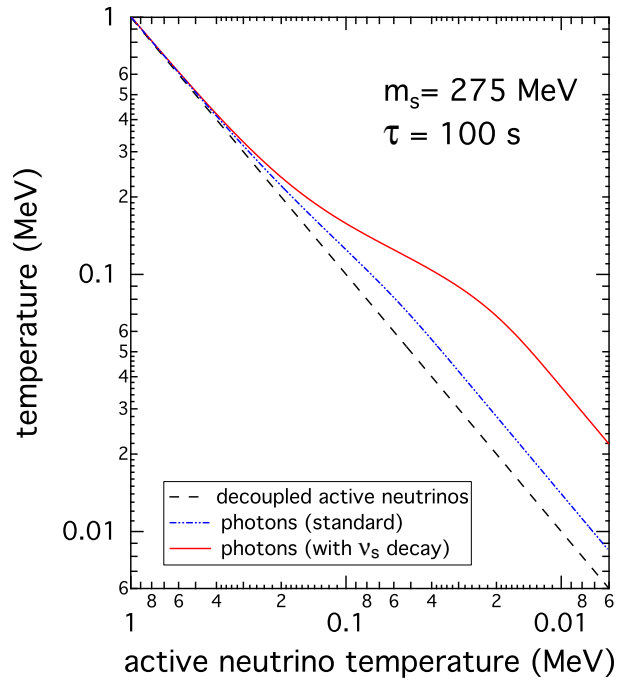}
\caption{Temperature versus decoupled active neutrino temperature for several scenarios as labeled. The decoupled active neutrino temperature (dashed line), decreasing to the left here, is inversely proportional to scale factor, which therefore increases to the right. The dash-dot-dot (blue) line shows the standard expansion with no sterile neutrinos, exhibiting the transfer of entropy from electron/positron pairs to photons as the former annihilate. The lighter solid (red) curve shows what happens in a scenario with a sterile neutrino with rest mass $m_s=275\,{\rm MeV}$ and lifetime $\tau=100\,{\rm s}$.}
\label{Temp}
\end{figure}

\begin{figure}
\includegraphics[width=3.7in]{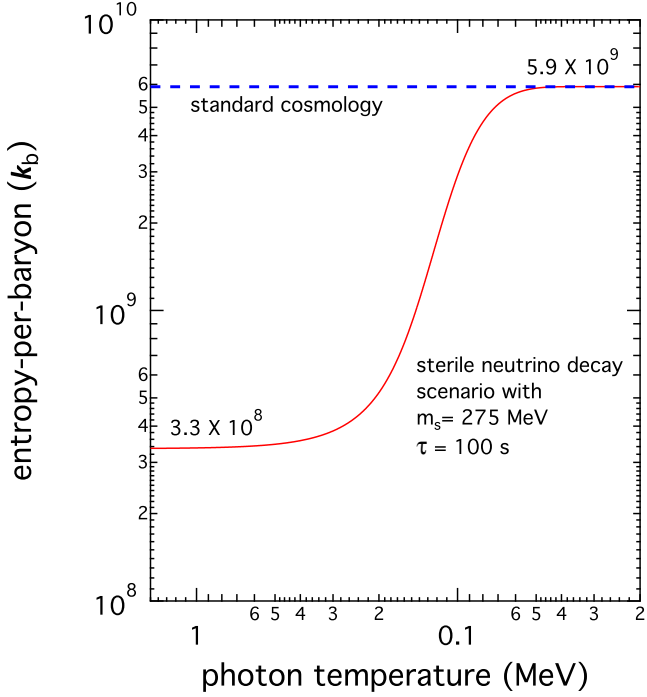}
\caption{Entropy-per-baryon (in units of Boltzmann's constant $k_{\rm b}$) versus photon (plasma) temperature for the standard cosmology case (constant co-moving entropy, dashed line) and for a scenario with a sterile neutrino with rest mass $m_s=275\,{\rm MeV}$ and lifetime $\tau=100\,{\rm s}$ (solid line). Beginning and ending entropy-per-baryon for the latter case as indicated.}
\label{entropy}
\end{figure}

Sterile neutrino decay adds entropy and so decreases the baryon-to-photon ratio, but in the end, at temperatures far below those characteristic of the BBN epoch, we must get the CMB-determined values of these quantities, {\it e.g.,} $\eta_{\rm WMAP} \approx 6.11\times{10}^{-10}$ as measured by WMAP \cite{Komatsu:2011lr} (used to scale Eq.~\ref{entropy1} above), implying $s_{\rm WMAP} \approx 5.895\times{10}^9$. Therefore, scenarios in which entropy is added through particle decay must start out with a higher value of $\eta_0$ (lower value of $s_0$).

We have solved Eq.~\ref{sdot} with a modified early universe expansion and Big Bang Nucleosynthesis code which treats all thermodynamic variables self consistently with the Friedman equation and all relevant weak interaction and sterile neutrino decay processes. 

Figure~\ref{Temp} shows the temperature of the plasma and the decoupled active neutrino seas as functions of decoupled active neutrino temperature, for several different scenarios, as calculated with our code. One scenario shown in this figure is just the standard radiation-dominated case, where $e^\pm$ annihilation occurs as electromagnetic equilibrium shifts with expansion, transferring entropy to the photons, but not to the active neutrinos. This results in the apparent shallowing of the slope of the plasma (photon) temperature near $T\sim 100\,{\rm keV}$. Entropy generated by sterile neutrino decay results in a similar, albeit more dramatic, phenomenon as is evident in Fig.~\ref{Temp}, where we show a scenario with particular sterile neutrino rest mass $m_s=275\,{\rm MeV}$ and lifetime $\tau=100\,{\rm s}$.

%
%

\begin{figure}
\includegraphics[width=3.7in]{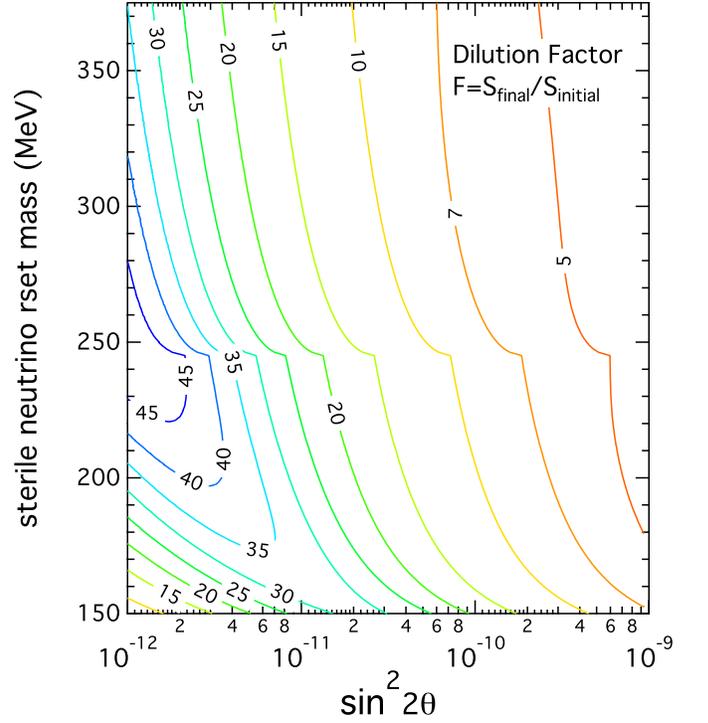}
\caption{Contours (as labeled) of dilution factor F (the ratio of final-to-initial entropy-per-baryon)
are given as functions of sterile neutrino rest mass in MeV and $\sin^22\theta$, where $\theta$ is the characteristic effective two-by-two vacuum mixing angle between active neutrino species and the sterile species.}
\label{F}
\end{figure}

\begin{figure}
\includegraphics[width=3.7in]{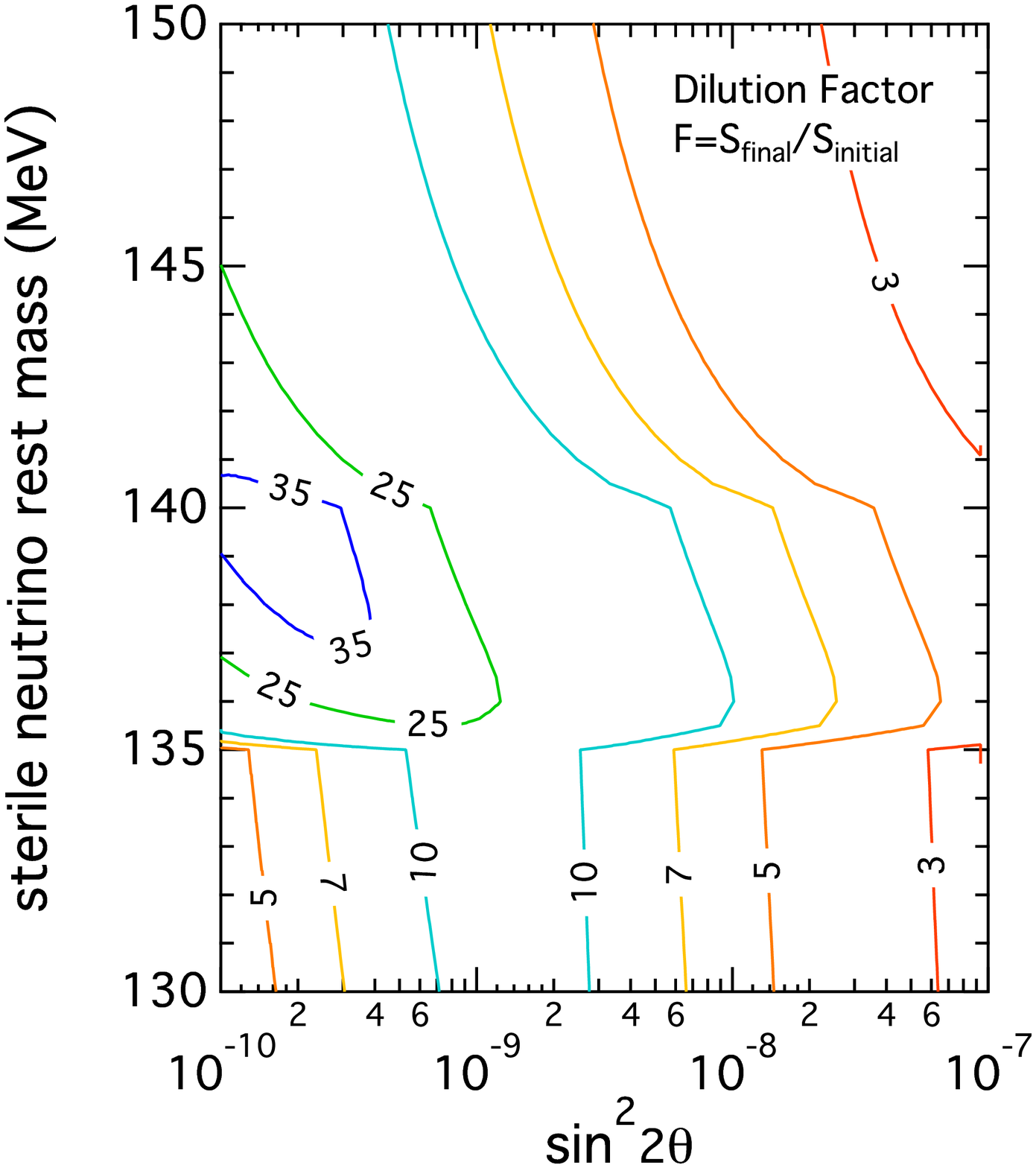}
\caption{Contours (as labeled) of dilution factor F (the ratio of final-to-initial entropy-per-baryon)
are given as functions of sterile neutrino rest mass in MeV and $\sin^22\theta$, where $\theta$ is the characteristic effective two-by-two vacuum mixing angle between active neutrino species and the sterile species.}
\label{F-L}
\end{figure}

\subsection{Dilution}

It is evident in the sterile neutrino decay scenario depicted in Fig.~\ref{Temp} that the fossil thermal neutrino relic background winds up considerably colder relative to the photons than in the standard cosmology case. This is dilution, a direct result of entropy generation, and analogous to what happens in the post-inflation re-heating epoch. In the cases we consider here the entropy is carried by relativistic particles. The ratio of the after-to-before entropies-per-baryon we designate as the dilution factor
\begin{equation}
F\equiv {{S_{\rm final}}\over{S_{\rm initial}}}= {{g_{s \rm f}\, T_{\rm f}^3\, a_{\rm f}^3  }\over{ g_{s \rm i}\, T_{\rm i}^3\, a_{\rm i}^3   }},
\label{Feqn}
\end{equation}  
where the indices i and f indicate initial (prior to any sterile neutrino decay) and final (long after sterile neutrino decay) values of statistical weight in relativistic particles $g_s$, photon temperature $T$, and scale factor $a$, respectively. Standard cosmology, with a fixed co-moving entropy, will have $F=1$. The relationship between scale factor and temperature for decoupled particles ({\it e.g.,} neutrinos) is $T_{\nu {\rm f}}\,a_{\rm f} = T_{\nu {\rm i}}\,a_{\rm i}$, and we assume that the initial photon and neutrino temperatures are the same, so that the ratio of the final neutrino temperature to the final photon temperature is
\begin{eqnarray}
{{ T_{\nu {\rm f}} }\over{ T_{{\rm f}} }} & = & {{1}\over{F^{\frac{1}{3}}}}\cdot{\left( {{ g_{s \rm f} }\over{g_{s \rm i} }}  \right)}^{\frac{1}{3}}\\
   & = &  {{1}\over{F^{\frac{1}{3}}}}\cdot{\left( {{2}\over{2+{\frac{7}{8}}\cdot 4 }}  \right)}^{\frac{1}{3}} = 
   {{1}\over{F^{\frac{1}{3}}}}\cdot{\left( {{4}\over{11 }}  \right)}^{\frac{1}{3}},
\label{tnu-t}
\end{eqnarray}
where in the last line we take as an example initial and final temperatures which bracket the BBN and $e^\pm$-annihilation epochs.
For the standard cosmology case, where $F=1$, we recover the usual relationship between the background neutrino temperature and the photon/plasma temperature at, {\it e.g.,} photon decoupling, ${{ T_{\nu {\rm f}} }/{ T_{{\rm f}} }} = {\left( {4}/{11} \right)}^{\frac{1}{3}} \approx 1/1.4$, {\it i.e.,} the final neutrino temperature should be $40\%$ lower than the photon temperature. But if entropy has been generated by out-of-equilibrium particle decay we will have $F > 1$, with a consequently lower ratio of background neutrino-to-photon temperatures. Therefore, for example, the ratio of the neutrino temperature in the diluted scenario at the photon decoupling epoch to the neutrino temperature at photon decoupling in the standard cosmology ($F=1$) case is
\begin{equation}
{{T_\nu^{\gamma\, {\rm dec}}}\over{T_\nu^{\gamma\, {\rm dec\ stan}}}} = F^{-1/3}.
\label{Tsbbn}
\end{equation}
The photon decoupling epoch is when the CMB photon temperature is $T\approx 0.2\,{\rm eV}$.

Figure~\ref{entropy} shows the history of entropy addition for a particular case (sterile neutrino rest mass $m_s = 275\,{\rm MeV}$ and lifetime $\tau = 100\,{\rm s}$) and for a standard constant co-moving entropy cosmology.  For this case $F=(5.9\times{10}^9)/(3.3\times{10}^8) \approx 18$, with $F^{\frac{1}{3}} \approx 2.6$, which implies a ratio of final neutrino and photon temperatures $2.6$ times lower than the standard case, roughly $1/3.6$. 

In general longer sterile neutrino decay lifetime for a given $m_s$ results in more entropy generation, necessitating lower starting entropy, hence higher starting $\eta_0$. For some choices of sterile neutrino parameters, sterile decay can be responsible for 90\% or more of $s_{\rm WMAP}$. Note that for the parameters chosen in the example in Fig.~\ref{entropy}, it is evident that the bulk of the entropy is added {\it during} the BBN epoch. This is true for a fair fraction of the sterile neutrino decay parameters considered here. BBN effects will be discussed in a subsequent paper.

We can use the code discussed above to survey cosmological effects for ranges of sterile neutrino masses and lifetimes. Dilution factors for a portions of these ranges are shown as contours in Fig~\ref{F} and Fig.~\ref{F-L}.
One trend in these figures is clear: the later the epoch at which the sterile neutrinos decay, the lower will be the temperature and, hence, the higher the ratio $m_s/T$ and so the higher the entropy added per sterile neutrino decay. For sterile neutrinos with masses $m_s < m_{\pi^0}$ three-neutrino decay will dominate the overall decay rate, implying relatively small $f$ and, hence, modest dilution. Once the sterile neutrino rest masses are larger than the $m_{\pi^0} \approx 135\,{\rm MeV}$ mass threshold, but smaller than the efficient decay neutrino-generating processes which operate above the $m_{\pi^\pm}+m_{e} \approx 140\,{\rm MeV}$ threshold, we see prodigious entropy generation and dilution. There is also substantial dilution {\it and} decay neutrino generation for longer lifetime sterile neutrinos with masses above the $m_{\pi^\pm}+m_{\mu} \approx 245.23\,{\rm MeV}$ threshold.

In Fig.~\ref{F-L} we see that the largest amounts of dilution correspond to ranges of sterile neutrino rest mass and vacuum mixing that fall in the \lq\lq sweet spot\rq\rq\ ranges which in some particle physics models \cite{Smirnov:2006uq} explain the observed light neutrino mass scale. This is the second regime of $m_s \,\theta^2$ discussed in section II and, as we shall see below, this range also may give an acceptable $N_{\rm eff}$.

\section{Generation of $N_{\rm eff}$ and an Altered Relic Neutrino Background}

Armed with the dilution factors and the average active neutrino energy per sterile neutrino decay, $\langle E_\nu^{\rm dec}\left( t\right)\rangle$, injected into the decoupled seas of active neutrinos at time $t$, we can calculate the radiation energy density and, hence, $N_{\rm eff}$ at the photon decoupling epoch. If at time $t$ the thermal background neutrino and sterile neutrino temperatures are $T_\nu\left( t\right)$ and $T_{\nu_s}\left( t\right)$, respectively, and $T_\nu^{\gamma\,{\rm dec}}$ is the thermal neutrino background temperature at photon decoupling, then we can use the definition of $N_{\rm eff}$ in Eq.~\ref{Neffdef} to show that in time step $\Delta t$ the increment to $N_{\rm eff}$ is  
\begin{widetext} 
\begin{eqnarray}
\Delta N_{\rm eff}\left( t\right) & = & {{   \langle E_\nu^{\rm dec}\left( t\right)\rangle  \cdot \left( \frac{3}{2} {{ \zeta\left( 3\right)}\over{\pi^2}}\, T_{\nu_s}^3\left( t\right)\ e^{-t/\tau} \ {{\Delta t}\over{\tau}} \right)   \cdot    {\left({{ T_\nu^{\gamma\,{\rm dec}}}\over{ T_\nu\left( t\right)}}  \right)}^4 }\over{   {{ 7 \pi^2 }\over{ 120}} \ {\left( T_\nu^{\gamma\, {\rm dec\, stan}}  \right)}^4   }} \\
& = & {{120 }\over{7 \pi^2 }} \cdot
{\left(  {{\langle E_\nu^{\rm dec}\left( t\right)\rangle}\over {T_\nu\left( t\right) }} \right)}  \cdot \left( \frac{3}{2} {{ \zeta\left( 3\right)}\over{\pi^2}} \ e^{-t/\tau} \ {{\Delta t}\over{\tau}} \right)   \cdot  {\left({{ T_{\nu_s}\left( t\right)}\over{ T_\nu\left( t\right)}}  \right)}^3 \cdot   {\left({{ T_\nu^{\gamma\,{\rm dec}}}\over{T_\nu^{\gamma\, {\rm dec\, stan}} }}  \right)}^4.
\label{DeltaNeff}
\end{eqnarray}
\end{widetext}
The first line in this equation is written in a way designed to elucidate its physical meaning. The numerator in this expression is a product of the neutrino energy injected per sterile neutrino decay and the second term, the number of decays per unit volume in $\Delta t$ (number density of sterile species at time $t$ multiplied by decay rate $1/\tau$ and $\Delta t$), and a third term which is simply a redshift factor (energy density redshifts like four inverse powers of scale factor and scale factor is inversely proportional to decoupled neutrino temperature). This numerator is therefore the energy density in active decay neutrinos produced by sterile neutrino decay at an earlier epoch $t$, redshifted down to the photon decoupling epoch. The factor in the denominator follows from the definition of $N_{\rm eff}$ in Eq.~\ref{Neffdef}.

\begin{figure}
\includegraphics[width=3.7in]{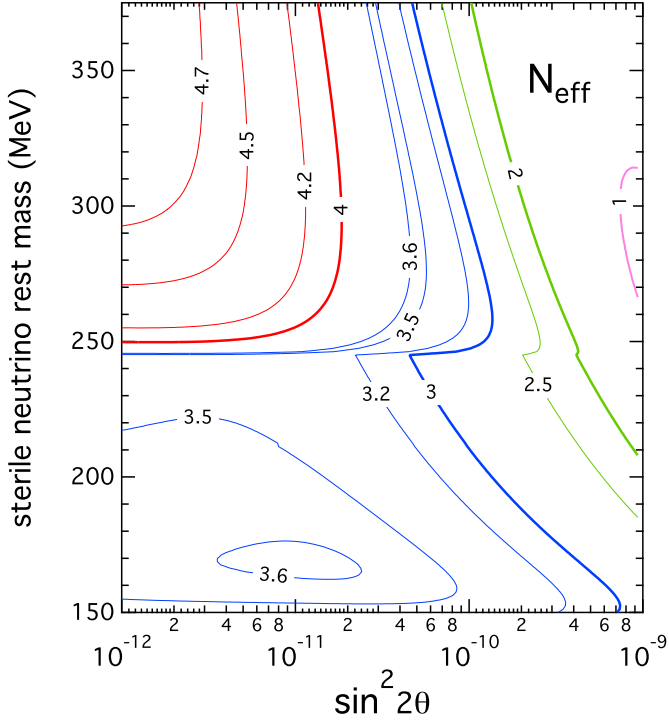}
\caption{Contours of ${\rm N}_{\rm eff}$ as functions of sterile neutrino rest mass (in MeV), and $\sin^22\theta$ as in Fig.~\ref{F}.}
\label{Neff}
\end{figure}

\begin{figure}
\includegraphics[width=3.7in]{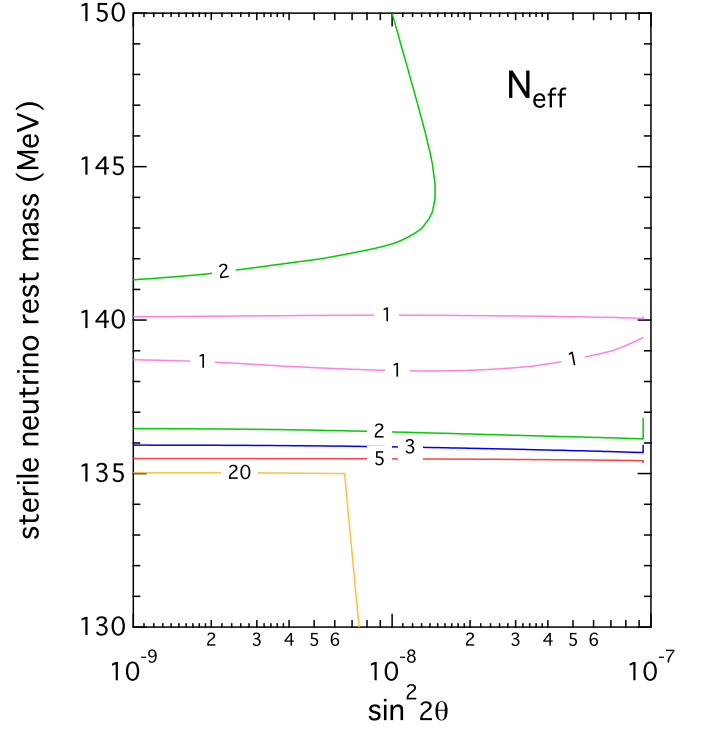}
\caption{Contours of ${\rm N}_{\rm eff}$ as functions of sterile neutrino rest mass (in MeV), and $\sin^22\theta$ as in Fig.~\ref{Neff}, but for a lower mass and larger vacuum mixing angle range.}
\label{Neff-L}
\end{figure} 

The second line, Eq.~\ref{DeltaNeff}, simply rearranges the terms in the first line to isolate co-moving invariants $\epsilon_{\rm dec}\left( t\right) \equiv {{\langle E_\nu^{\rm dec}\left( t\right)\rangle}/ {T_\nu\left( t\right) }}$ and $G= {\left({{ T_{\nu_s}\left( t\right)}/{ T_\nu\left( t\right)}}  \right)}$. Using these co-moving quantities and Eq.~\ref{Tsbbn} in Eq.~\ref{DeltaNeff}, we can conclude that 
\begin{equation}
{{d N_{\rm eff}}\over{d t}}={\left[ {{{ 180\,\zeta\left( 3\right) }\over{7 \pi^4 }} }\right]} \cdot {{G^3 }\over{F^{\frac{4}{3}} }} \cdot \epsilon_{\rm dec}\left( t\right) \cdot {{e^{-t/\tau}}\over{\tau}}.
\label{de}
\end{equation}
We can use these considerations in the code described above to integrate this equation and derive $N_{\rm eff}$ as a function of sterile neutrino rest mass and lifetime.

Fig.~\ref{Neff} and Fig.~\ref{Neff-L} show the results of these calculations. They give contours of $N_{\rm eff}$ as functions of sterile neutrino rest mass and $\sin^22\theta$, essentially in ranges of these parameters identical to those used in the lifetime and dilution factor figures above. These results are remarkable. Where dilution is large, yet not accompanied by significant active decay neutrino generation ({\it e.g.,} where $m_s$ is less than the charged pion decay channel thresholds, yet bigger than the $\pi^0$ threshold), $N_{\rm eff}$ is smaller than $3$, reflecting the dilution-caused reduction/refrigeration of the ordinary thermal neutrino background contribution to the radiation energy density at photon decoupling. On the other hand, for $m_s$ large enough for sterile neutrino decay to proceed through $\nu_s\rightarrow \pi^\pm+\mu^\mp$, we will have prodigious active decay neutrino production {\it and} dilution-caused reduction of the the ordinary thermal background neutrino contribution to $N_{\rm eff}$. However, as is evident in Fig.~\ref{Neff}, the former effect wins out over dilution and large values of $N_{\rm eff}$ can be generated, reflecting the large amount of decay neutrino energy injected into the active neutrino seas during the epoch of sterile neutrino decay in this case.

The particular scenario with $m_s=275\,{\rm MeV}$ and $\tau= 100\,{\rm s}$ shown in Fig.~\ref{Temp} and Fig.~\ref{entropy} serves to illustrate these points. This case has dilution factor $F=17.6$ and results in $N_{\rm eff} = 4.266$, and of that only $N_{\rm eff}^{\rm therm} = 0.0655$ is contributed by the ordinary thermal background neutrinos. (This small contribution from the thermal background neutrinos follows from dilution: $N_{\rm eff}^{\rm therm} \approx 3\cdot F^{-4/3}\approx 3\cdot {\left( 17.6 \right)}^{-4/3} \approx 0.0655$.) 

It is a disturbing thought that there could be ranges of sterile neutrino rest mass and lifetime, unconstrained by laboratory experiments and existing cosmological bounds, with $N_{\rm eff} \approx 3$ in seeming accord with standard cosmology, but where the ordinary thermal background neutrinos are grossly diluted, comprising only $\sim 2\%$ of $N_{\rm eff}$. Perhaps BBN considerations will rule out these sterile neutrino parameters, but we already have constraints on $N_{\rm eff}$ from CMB considerations (WMAP7 \cite{Komatsu:2011lr} reports $N_{\rm eff} = 4.34^{+0.86}_{-0.88}$) which promise to get much better with future observations. The current constraints, while not very good, already serve to eliminate much of the sterile neutrino parameter space depicted in Fig.~\ref{Neff} and Fig.~\ref{Neff-L}, especially for the higher and lower sterile neutrino rest masses on these figures and especially for relatively longer lifetimes (lower effective vacuum active-sterile mixing angles) on these plots. 


\begin{widetext}

\begin{figure}
\includegraphics[width=3.7in, angle=270]{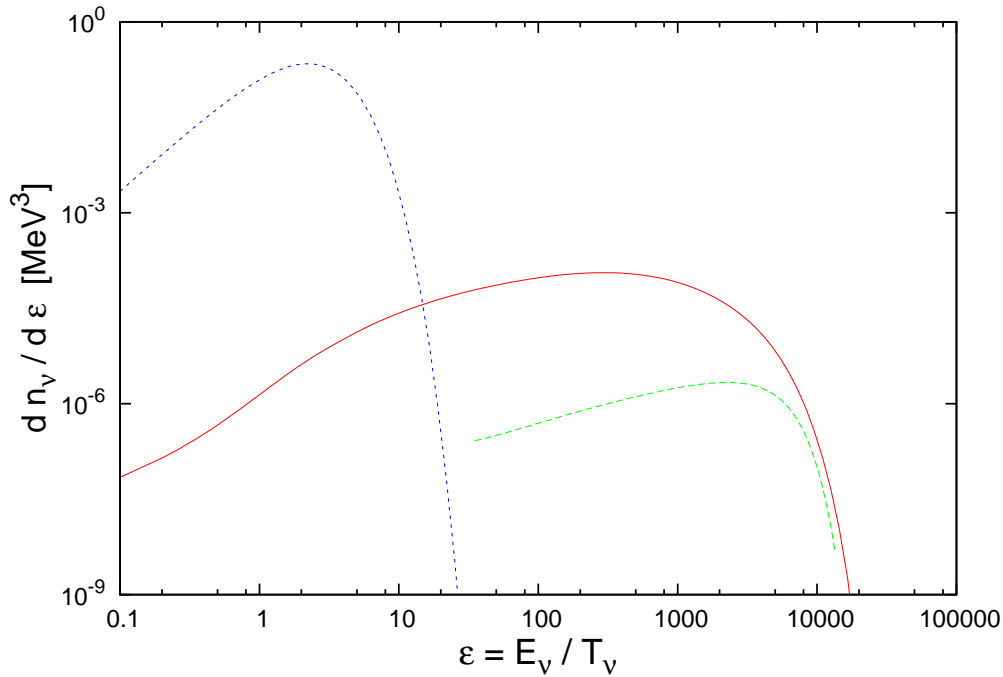}
\caption{The relic neutrino energy spectrum for the particular case with $m_s=275\,{\rm MeV}$ and $\tau=100\,{\rm s}$. Here the neutrino number density per co-moving scaled neutrino energy (${d n_\nu}/{d \epsilon}$ in ${\rm MeV}^3$) is shown as a function of the dimensionless co-moving scaled neutrino energy parameter $\epsilon=E_\nu/T_\nu$ for three components: the ordinary thermal background neutrinos (short-dashed, blue line); sterile neutrino decay-generated neutrinos from the charged pion decay channels (solid, red line), and the $\pi^0$ decay channel (dashed, green line).}
\label{nuspectrum}
\end{figure}
\end{widetext}

Given the spectral energy distributions of the active neutrinos resulting from the various sterile neutrino decay channels \cite{Fetscher:1992lr,Greub:1994fk}, we can use the above considerations and calculate the actual spectrum of active neutrinos left over from sterile neutrino decay. Fig.~\ref{nuspectrum} shows the the relic neutrino energy spectrum for the particular case with $m_s=275\,{\rm MeV}$ and $\tau=100\,{\rm s}$. This figure shows three components to the relic neutrino background in this scenario: the ordinary thermal background ({\it e.g.,} see Ref.s~\cite{Fuller:2009lr,Dodelson:2009qy}); and the decay-generated neutrinos from the charged pion decay channels and the $\pi^0$ decay channel. For the sterile neutrino lifetime in this case, most of the decays will be where the age of the universe is order $t  > 100\,{\rm s}$, where the average thermal neutrino energy will be $T_\nu < 100\,{\rm keV}$, but the average decay neutrino energy will be more like $50$ to $100\,{\rm MeV}$ in this case, roughly a thousand times the thermal neutrino energy. By the time of photon decoupling, the ratio of the thermal background neutrino temperature to the photon temperature in this case will be $T_\nu^{\gamma\,{\rm dec}}/T^{\gamma\,{\rm dec}} = (4/11)^{(1/3)} \cdot F^{-1/3} \approx (4/11)^{1/3} \cdot {(1/17.6)}^{1/3} \approx 0.27$. Since $T^{\gamma\,{\rm dec}} \approx 0.2\,{\rm eV}$ this implies that the average energy of the thermal background neutrinos will be $3\,T_{\nu}^{\gamma\,{\rm dec}} \sim 0.16\,{\rm eV}$, barely large enough to assure that these species are relativistic at this epoch. The minimum neutrino mass eigenvalue is the square root of the atmospheric neutrino mass-squared difference, $\sqrt{\delta m^2_{\rm atm}} \approx 0.055\,{\rm eV}$. (The spectrum shown in Fig.~\ref{nuspectrum} remains correct even if the kinematics of the neutrinos becomes non-relativistic, so long as $E_\nu$ is interpreted as the neutrino spacelike momentum magnitude.) Note that the peak of the decay neutrino relic spectrum is about a factor of $1000$ larger than that of the background thermal neutrinos. This implies that the decay neutrinos have energies $\sim 160\,{\rm eV}$ at the photon decoupling epoch. For a maximum neutrino mass eigenvalue $\sqrt{\delta m^2_{\rm atm}}$, these neutrinos would still be relativistic at the current epoch, as the ratio of $160/.055 \approx 2900$, but the redshift at photon decoupling is only $z_{\gamma\,{\rm dec}} \approx 1100$.
Note also that in this scenario the number density of neutrinos in the thermal background is only about $6\%$ ({\it i.e.,} $1/F \approx 0.06$) of the number density in a standard cosmology. Moreover, the total number density of decay neutrinos in the relic background is $\sim 10\%$ of that in the thermal neutrino relic background. 

The upshot of the number densities of the relic neutrinos being low compared to the standard cosmology case and the decay neutrinos being at very much higher energies than in this standard case is that it is unlikely that cosmological neutrino mass measurements will detect any neutrino mass in these sterile neutrino scenarios, even if $N_{\rm eff}$ is within observational bounds. 

In fact, a \lq\lq smoking gun\rq\rq\ signal for a dilution and particle decay scenario along the lines of that discussed here would be the measurement of a radiation energy density at photon decoupling with $N_{\rm eff} \neq 3$ and yet no detection of a neutrino mass even when the observational sensitivity to the sum of the light neutrino masses is pushed below known, laboratory mass limits, either $2  \sqrt{\delta m^2_{\rm atm}} \sim 100\,{\rm meV}$ in the inverted neutrino mass hierarchy, $\sqrt{\delta m^2_{\rm atm}}\sim 50\,{\rm meV}$ in the normal mass hierarchy, or below a mass measured directly by tritium endpoint experiments ({\it e.g.,} KATRIN) or neutrino-less double beta decay experiments. Prospects for bettering sensitivity to the sum of the neutrino masses in future observations and analyses are discussed in Ref.~\cite{Abazajian:2011fk}.


\section{Sterile Neutrino-Engendered Matter Dominated Epochs in the Early Universe}

For most of the ranges of sterile neutrino rest mass and decay lifetime discussed above there will be significant periods of matter domination in the early universe before, during, or even after the BBN epoch. This is not surprising because the sterile neutrinos are nearly as numerous as photons and they decay at temperature scales much lower than their rest masses. 

\begin{figure}
\includegraphics[width=3.7in]{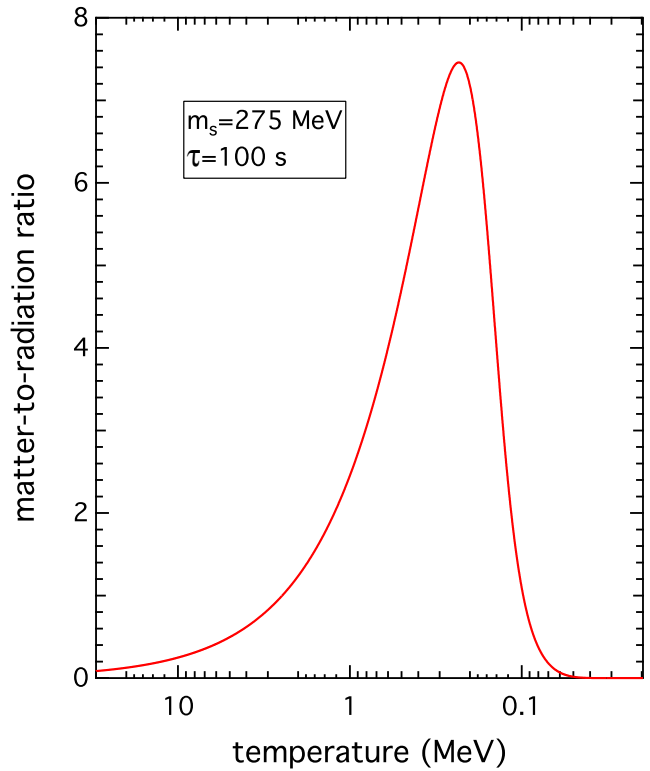}
\caption{Ratio of matter-to-radiation energy density as a function of plasma/photon temperature (in MeV) for the indicated sterile neutrino rest mass, $m_s$, and lifetime $\tau$.}
\label{matt-rad}
\end{figure}

Fig.~\ref{matt-rad} shows the ratio of the energy density in non-relativistic matter (sterile neutrino rest mass) as a function of photon temperature for the particular case with sterile neutrino rest mass $m_s = 275\,{\rm MeV}$ and lifetime $\tau = 100\,{\rm s}$. In this scenario we see that matter can dominate over radiation by nearly an order of magnitude for what turns out to be many Hubble times at the sterile neutrino decay epoch. The peak of entropy generation in this scenario is, of course, during the peak epoch of sterile neutrino decay, which will tend to be somewhat later than peak matter domination. Moreover, the scale-factor/time relation is considerably altered here.

In the particular case shown in Fig.~\ref{matt-rad} the matter-dominated epoch occurs just prior to, and in the early stages of BBN. In a standard BBN scenario a scale-factor/time relation characteristic of matter-domination  produces light element abundance yields which look nothing like the observationally-inferred primordial abundances, especially for $^4{\rm He}$. In these sterile neutrino decay scenarios, however, entropy is being added to the system and BBN remains close to nuclear statistical equilibrium for the lighter elements, so it is not a foregone conclusion that BBN considerations can rule out all of the parameter space of sterile neutrino mass and lifetime not already ruled out by observationally-derived $N_{\rm eff}$ constraints. We will discuss detailed nucleosynthesis calculations for these scenarios in a subsequent work.

\begin{figure}
\includegraphics[width=3.7in]{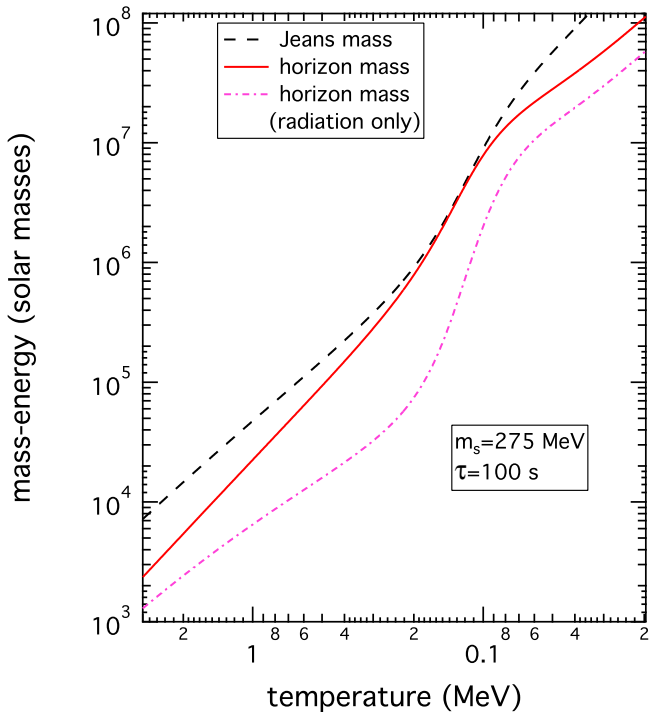}
\caption{Causal horizon mass (in solar masses) as a function of plasma/photon temperature (in MeV) for a scenario (solid line) with the indicated sterile neutrino rest mass, $m_s$, and lifetime $\tau$, and the horizon mass that would be inferred from the radiation energy density alone (dash-dotted line). The dashed line gives the Jeans mass for this sterile neutrino scenario.}
\label{a-T}
\end{figure}

Such drastic changes to the time-temperature-scale factor relationships that accompany sterile neutrino decay scenarios also result in altered  causal and dynamical effects in the early universe. In broad brush, more energy density from sterile neutrino rest mass results in faster expansion rates and larger causal horizon lengths at a given temperature. But matter-domination adds an additional twist, driving the total mass-energy inside the horizon toward the Jeans mass.  

Fig.~\ref{a-T} shows the total mass-energy inside the causal horizon as a function of photon temperature for the particular scenario with sterile neutrino rest mass $m_s = 275\,{\rm MeV}$ and lifetime $\tau = 100\,{\rm s}$. A causal horizon mass calculated using only the thermal radiation energy density is shown to illustrate the effect of sterile neutrino rest mass. The central take-awy message of this figure is that the mass scales affected by these sterile neutrino scenarios are small, of order the horizon scale $M_{\rm H} \sim {10}^6$ to ${10}^7\,{\rm M}_\odot$, or smaller. These scales are marginally below what can currently be probed with CMB anisotropy and large scale structure arguments.

This figure also shows the instantaneous Jeans mass as a function of temperature in the plasma. The Jeans mass $M_{\rm J} = c_s^3  {\left( \pi\,m_{\rm pl}^3 \right)}/{\left( 6\,\sqrt{\rho} \right)}$, where $c_s$ is the sound speed, is derived by setting the sound crossing time across a scale equal to the gravitational collapse (free fall) time on that scale. Generally, mass scales larger than the Jeans mass are subject to collapse, and fluctuations inside the horizon can grow in amplitude in matter-dominated conditions. In a radiation-dominated Friedman universe the causal horizon mass will be $M_{\rm H} = {\left( 3/8 \pi  \right)}^{3/2}\, (1/c_s^3)\,M_{\rm J}$, {\it i.e.,} less than the Jeans mass, and fluctuation growth is suppressed. 

In these matter-dominated sterile neutrino scenarios the sound speed is still set by radiation and remains close to light speed, $c_s \approx 1/\sqrt{3}\ c$. This means that, at best, the horizon mass and Jeans mass nearly coincide. This is evident in Fig.~\ref{a-T}. We conclude that collapse of horizon-sized regions is unlikely because there is no way for such structures to dump their entropy and cool. At best, the effect of the matter-dominated epoch engendered by sterile neutrinos might be to peak-up power on sub-horizon scales. These scales are relatively small.

\section{Implications and Conclusions}

We have assessed the impact of massive decaying sterile neutrinos on the thermal and dynamical history of the early universe and we have used our calculations to make predictions of fossil radiation energy density in these scenarios which can be used to constrain them.  In particular, we have examined sterile neutrinos with rest masses in the $100\,{\rm MeV}$ to $500\,{\rm MeV}$ range with vacuum coupling/mixing with ordinary active neutrinos large enough to engender thermal and chemical equilibrium for these sterile species at very early epochs in the history of the universe ({\it i.e.,} $T > 1\,{\rm GeV}$), but small enough that they decay with lifetimes of seconds to minutes. 

We have shown that the out-of-equilibrium decay of these sterile species can lead to prodigious entropy production, as well as the generation of a significant population of decoupled active neutrinos. In these scenarios $90\%$ or more of the observed entropy-per-baryon in the universe can be generated by these decays. This necessarily means that pre-existing decoupled background particles, {\it e.g.,} the thermal background neutrinos, will suffer dilution, {\it i.e.,} reduction in their number densities and energies beyond that produced in a standard cosmological expansion.

This can engineer a seemingly bizarre result. We find, for example, that ranges of sterile neutrino rest masses and lifetimes could produce a radiation energy density (as parameterized by $N_{\rm eff}$) at the photon decoupling epoch within current observational bounds, but where the ordinary thermal background neutrinos make only a negligible contribution to this.

This dilution has a robust signature: a measured $N_{\rm eff} \neq 3$ (either smaller than $3$ or significantly larger than this) coupled with {\it no} cosmological signal for neutrino rest mass, even when the detection thresholds for the sum of the light neutrino masses in these probes are below laboratory-measured neutrino mass values, either as established by the atmospheric neutrino oscillation scale ($\sim 50\,{\rm meV}$ in the normal neutrino mass hierarchy, $\sim 100\,{\rm meV}$ in the inverted) or direct measurements with, {\it e.g.,} tritium endpoint experiments like KATRIN or neutrino-less double beta decay experiments. There would be no neutrino mass detected in these scenarios because cosmological probes are predicated on the assumption of black body, Fermi-Dirac-shaped spectral energy distributions (with temperature parameter some $40\%$ less than the photon background temperature) for the relic background neutrinos and this would not be the case with the significant dilution attending the sterile neutrino models considered here.

Of course, many of the sterile neutrino rest mass and lifetime parameters considered here will affect light element abundance yields in Big Bang Nucleosynthesis (BBN). Likely this can result in additional constraints beyond the energy density constraints presented here. What, however, would be the implications of sterile neutrino with masses and lifetimes which somehow managed to evade these constraints?

The answer is straightforward if somewhat unsettling. We have shown that fluctuations on scales of $\sim {10}^7\,{\rm M}_\odot$ could be affected, albeit seemingly minimally. More significantly, the massive dilution which is a by-product of sterile neutrino decay in these scenarios has two effects. First, since in these sterile neutrino scenarios there can be more than an order of magnitude increase in its baryon-to-entropy ratio, the pressure on existing baryogenesis models would be significantly increased. This is especially true for electroweak baryogenesis schemes \cite{Kuzmin:1985lr,Rubakov:1996fk,Cirigliano:2010yq,Cirigliano:2011fj}. 

Second, dilution can hide many sins. Light decoupled particles could be \lq\lq hidden\rq\rq\ from cosmological detection and constraint by dilution. For example, light sterile neutrinos,  possibly those suggested by recent experiments \cite{Athanassopoulos:1995vn,Athanassopoulos:1996yq,Athanassopoulos:1998kx,Athanassopoulos:1998fj,Aguilar:2001fj,Adamson:2009uq,Aguilar-Arevalo:2009qy,Aguilar-Arevalo:2010fk,Abe:2011uq,MiniB:2011kx,MiniB:2011lr} and some of those long popular with supernova heavy element nucleosynthesis modelers \cite{McLaughlin:1999fk}, but seemingly ruled out by cosmological bounds, now might be compatible with these bounds.
The dilution schemes discussed here might then be an alternative to attempts to reconcile evidence for a light sterile with existing bounds \cite{Hamann:2010lr,Hamann:2011lr, Hamann:2011fk}. 

There are caveats. Though dilution could reduce the number densities and energies of these particles enough to evade existing cosmological bounds, this is only strictly true if their abundances are set prior to the entropy generation/dilution epoch. If these light sterile species had large enough vacuum mixing angles then oscillations could regenerate appreciable populations of these particles from the relic backgrounds left after dilution, and then the cosmological constraints might apply again if the relic decay neutrinos and thermal background neutrinos had a high enough number density post entropy generation \-- which they may not have for enough dilution and low enough light sterile neutrino mass. Obviously BBN constraints on dilution scenarios can be another severe constraint, as it is not yet settled whether any of the dilution schemes discussed here can be reconciled with light element abundance bounds.

Curiously, some of the ranges of sterile neutrino rest mass and vacuum mixing with active species which we have found to give acceptable values of $N_{\rm eff}$, and considerable dilution, also fall in the ranges of the parameters required by models to explain the light neutrino masses \cite{Smirnov:2006uq}. 

Of late we have learned much from the laboratory about the properties of neutrinos. However, the now established fact that neutrinos have rest masses opens a Pandora's Box of possibilities for new physics in the neutrino sector \cite{Mohapatra:2006qy,Smirnov:2006uq,Mohapatra:2007lr,de-Gouvea:2009fk}. These possibilities, while highlighting our ignorance, nevertheless may make for a future golden age of neutrino physics when the next generation cosmological probes \cite{Hamann:2008pd,Galli:2010ul,Shimon:2010qf,Benetti:2011mz} become available.

\begin{acknowledgments}
We would like to acknowledge helpful discussions with J. Carlson, A. Friedland, E. Grohs, A. Hayes, W. C. Haxton, W. Louis, K. Petraki, and C. J. Smith.  This work was supported in part by NSF Grant No. PHY-09-70064 at UCSD and DOE grant DE-FG03-91ER40662 at UCLA.
\end{acknowledgments}

\bibliography{/Users/gfuller/XBibTeX-references/allref}

\end{document}